\documentclass[aps,prmaterials,preprint,superscriptaddress]{revtex4-2}
\usepackage{gensymb}
\usepackage{graphicx}
\usepackage{booktabs}
\usepackage{multirow}
\usepackage{chemformula}
\usepackage{appendix}
\begin{document}
\title{High-pressure synthesis of quantum magnet $M$-\ch{YbTaO4} with a stretched diamond lattice}
\author{Nicola D.~Kelly}
\email[]{ne281@cam.ac.uk}
\affiliation{Cavendish Laboratory, University of Cambridge, J J Thomson Avenue, Cambridge, CB3 0US, United Kingdom}
\affiliation{Research Center for Materials Nanoarchitectonics (MANA), National Institute for Materials Science (NIMS), 1-1 Namiki, Tsukuba, Ibaraki 305-0044, Japan}

\author{Xuan Liang}
\affiliation{Research Center for Materials Nanoarchitectonics (MANA), National Institute for Materials Science (NIMS), 1-1 Namiki, Tsukuba, Ibaraki 305-0044, Japan}

\author{Si\^{a}n E.~Dutton}
\email[]{sed33@cam.ac.uk}
\affiliation{Cavendish Laboratory, University of Cambridge, J J Thomson Avenue, Cambridge, CB3 0US, United Kingdom}

\author{Kazunari Yamaura}
\affiliation{Research Center for Materials Nanoarchitectonics (MANA), National Institute for Materials Science (NIMS), 1-1 Namiki, Tsukuba, Ibaraki 305-0044, Japan}

\author{Yoshihiro Tsujimoto}
\email[]{tsujimoto.yoshihiro@nims.go.jp}
\affiliation{Research Center for Materials Nanoarchitectonics (MANA), National Institute for Materials Science (NIMS), 1-1 Namiki, Tsukuba, Ibaraki 305-0044, Japan}


\begin{abstract}
We report bulk magnetic properties of ytterbium tantalate in its monoclinic fergusonite modification, $M$-\ch{YbTaO4}. The spin-$\frac{1}{2}$ Yb$^{3+}$ ions in this phase are arranged on a geometrically frustrated ``stretched diamond" lattice. $M$-\ch{YbTaO4} cannot be prepared at ambient pressure and was instead prepared in a belt-type apparatus at 6~GPa and 1800~\degree C. Susceptibility and specific heat data show no long-range ordering down to 1.8~K and are consistent with a $J_\mathrm{eff}=\frac{1}{2}$ Kramers doublet which splits in an applied field. Furthermore, under high-pressure synthesis the entire solid solution \ch{YbNb_{x}Ta_{1-x}O4} ($0\leq x \leq 1$) can be stabilised in the $M$ phase, in contrast to ambient-pressure synthesis which favours the competing $M'$ phase for Ta-rich compositions. Subsequent annealing of the Nb-Ta mixed samples resulted in colour changes, suggesting oxygen deficiency in some of the as-prepared high pressure samples. There was little variation in the bulk magnetic properties upon varying either the Nb/Ta ratio or the annealing conditions.
\end{abstract}
\maketitle

\section{Introduction}

Synthesis under high pressure is an important tool to stabilise novel phases of condensed matter. It has been used to synthesise compounds from many different chemical and structural families, including borates \cite{Huppertz2001,Huppertz2002,Emme2004}, oxychalcogenides \cite{Song2022,Takeiri2016} and oxyhalides \cite{Tsujimoto2022}, nitrides \cite{Yuan2024}, simple, double, triple and quadruple perovskite oxides \cite{Markkula2011,Almadhi2024,Liu2025,Solana-Madruga2021,Liang2026}, and many others, often with unusual magnetic or electronic properties such as incommensurate magnetic order. High-pressure-high-temperature synthesis is a particularly powerful technique to stabilise cations in unusual oxidation states or coordination environments \cite{Solana-Madruga2021}, allowing the synthesis of metastable phases that would not be accessible at ambient pressure.

Solid-state materials with the magnetic ions located on a diamond-like lattice have gained significant attention in recent years. In a perfect cubic diamond system, such as the $A$-sites of spinels like \ch{MnAl2O4} \cite{Tristan2005} or \ch{CoRh2O4} \cite{Ge2017}, all the nearest-neighbour interactions between magnetic ions are identical, forming a network where every magnetic ion is surrounded by four others in a tetrahedral arrangement. Considering only the nearest-neighbour exchange interactions ($J_1$), spins on a cubic diamond lattice are expected to display long-range three-dimensional order. Both \ch{MnAl2O4} and \ch{CoRh2O4} are antiferromagnets with N\'{e}el temperatures of 40 and 25~K respectively. However, non-zero next-nearest-neighbour interactions $J_2$ can also compete with $J_1$, sometimes significantly suppressing long-range order \cite{Fritsch2004} or at other times promoting a glassy state \cite{Cho2020,Tristan2005}, showing that the cubic diamond lattice is already a rich landscape for different magnetic behaviours. In contrast, in tetragonal or hexagonal crystal structures such as those of \ch{LiYbO2} \cite{Bordelon2021a} or $\beta$-\ch{KTi(C2O4)2* 2 H2O} \cite{Abdeldaim2020}, the distances remain identical but the angles between adjacent spin-spin interactions are distorted away from the ideal 109.5\degree, leading to competition between different interactions. Especially when there are non-zero $J_2$ interactions, such a network can also host strong geometric magnetic frustration \cite{Marjerrison2016,Chamorro2018,Ge2017,Bordelon2021}. Further distortion or ``stretching" of the diamond lattice occurs in monoclinic systems, such as the lanthanide metaborates \ch{\textit{Ln}(BO2)3} \cite{Mukherjee2017}, niobates \ch{\textit{Ln}NbO4} and tantalates $M$-\ch{\textit{Ln}TaO4} \cite{Kelly2022a}. Frustration effects in compounds with a stretched diamond lattice have been linked to observations of unusual magnetic behaviour including incommensurate helical order \cite{Bordelon2021}. Furthermore, strong spin-orbit coupling and crystal electric field effects also play an important part in the magnetism of lanthanide-based compounds.

A recent study focused on the series of compounds $M$-\ch{\textit{Ln}TaO4} ($Ln=$ Nd--Er) with a monoclinic stretched diamond lattice of magnetic $Ln^{3+}$ ions. None of the compounds orders above 1.8~K except $M$-\ch{TbTaO4}, which displays $A$-type antiferromagnetic (AFM) order below $T_\mathrm{N}=2.25$~K \cite{Kelly2022a}. In an applied field the compound undergoes a spin-flop-type transition to ferromagnetic (FM) at 6~T and 1.6~K and there is evidence for magnetoelectric coupling \cite{Zhang2025}. Furthermore, Kumar \textit{et al.} have recently studied the isostructural ytterbium niobate \ch{YbNbO4} using magnetic susceptibility, heat capacity and muon-spin relaxation ($\mu$SR) spectroscopy. Their observations exclude the possibility of either long-range order or spin freezing and indicate that the Yb$^{3+}$ spins remain correlated and strongly fluctuating down to 300~mK, making \ch{YbNbO4} a potential spin liquid candidate \cite{Kumar2025}. 

\begin{figure}[htbp]
\centering
\includegraphics[width=0.5\textwidth]{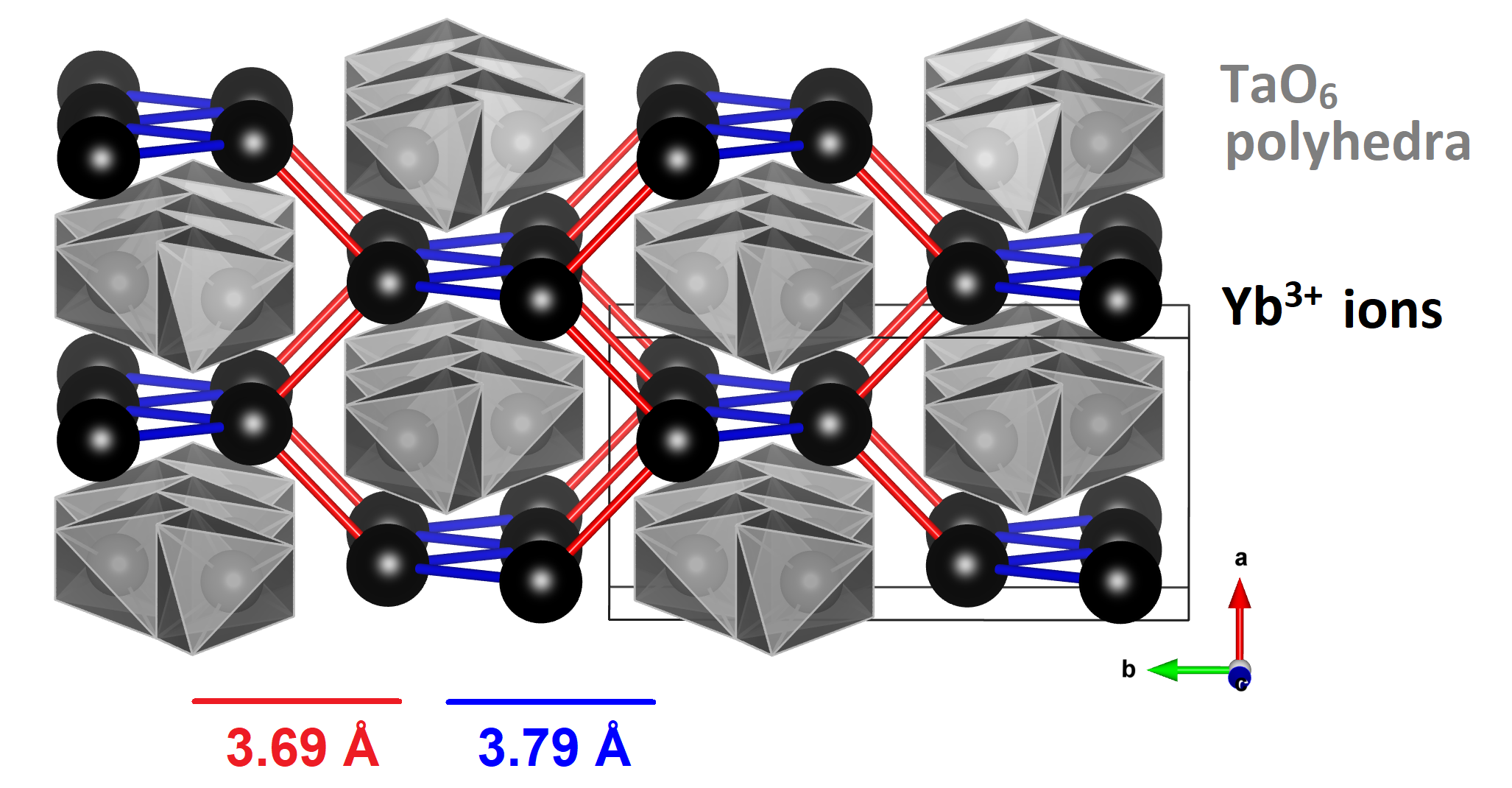}
\caption{Crystal structure of $M$-\ch{YbTaO4}, space group $I2/a$ (No.~15). Yb$^{3+}$ ions are shown in black and Ta$^{5+}$ ions in grey; oxide ions are located at the vertices of the \ch{TaO6} polyhedra. The red and blue solid lines indicate the two closest Yb--Yb distances making up the stretched diamond lattice.}
\label{fig:diamond}
\end{figure}

The $M$-phase of \ch{YbTaO4} is isostructural with \ch{YbNbO4} and therefore also hosts a stretched diamond lattice, Fig.~\ref{fig:diamond}. The presence of two competing Yb--Yb interactions with similar interatomic distances is expected to produce geometric magnetic frustration. It is not possible to prepare $M$-\ch{YbTaO4} at ambient pressure and all previous reports \cite{Brixner1983,Markiv2002a} have utilised high-pressure synthesis. To our knowledge no magnetic studies have been carried out on this phase. In this work, we synthesised $M$-\ch{YbTaO4} with high reproducibility using a belt-type high-pressure apparatus at 1800~\degree C and 6~GPa and measured its bulk magnetic properties down to 1.8~K. $M$-\ch{YbTaO4} shows no long-range order in this temperature range, and magnetometry and specific heat data support an effective $J=\frac{1}{2}$ Kramers doublet state. We also synthesised the solid solution \ch{YbNb_{x}Ta_{1-x}O4} ($0\leq x\leq 1$) under both ambient- and high-pressure conditions. Under ambient pressure, Nb-rich samples crystallise in the $M$ phase with a stretched diamond lattice whilst Ta-rich samples favour the competing $M'$ phase with a distorted 2D square lattice of Yb$^{3+}$ ions \cite{Kumar2024,Ramanathan2024}. In contrast, high-pressure synthesis allows the pure $M$-phase to be stabilised for all $x$. However, the reducing environment of the high-pressure synthesis resulted in coloured samples for some values of $x$, which was ascribed to oxygen deficiency; the colour changed to white after annealing under ambient pressure in either air or \ch{O2}.

\section{Experimental Methods}

\subsection{Synthesis}
Appropriate stoichiometric amounts of \ch{Yb2O3} (99.9\%, Kojundo Chemical Laboratory Co. Ltd), \ch{Nb2O5} (99.9\%, Rare Metallic Co. Ltd) and \ch{Ta2O5} (99.9\%, Rare Metallic Co. Ltd) for a target mass of 0.5~g \ch{YbNb_{x}Ta_{1-x}O4} ($0\leq x\leq1$) were weighed accurately and ground together by hand. For ambient-pressure synthesis, the mixture was pressed into an 8~mm pellet, placed in an alumina crucible and heated at 3~\degree C~min$^{-1}$ to 1500~\degree C for 24~h in air. After cooling at the natural rate of the furnace, the resultant pellets were ground into fine white powders. For high-pressure synthesis, the mixture was sealed into a Pt capsule (inner diameter 6.9~mm, length $\approx2.2$~mm, wall thickness $\approx0.2$~mm) and loaded into a belt-type press (Kobe Steel, Ltd.) \cite{TracyHall1960} using a NaCl–pyrophyllite cell assembly with a graphite tube heater \cite{Miyakawa2015}. After cold-pressurizing to 6 GPa, the samples were heated to 1800~\degree C at 100~\degree\ min$^{-1}$, held for 60 min, and quenched to below 100~\degree C within 30~s while holding pressure. Pressure was then released over approximately 40 minutes and the product was recovered from the Pt capsules in the form of a white ($x=0$) or beige ($x>0$) fine powder. A sample of $M$-\ch{LuTaO4} was synthesised from \ch{Lu2O3} (99.9\%, Kojundo Chemical Laboratory Co. Ltd) and \ch{Ta2O5} at 6~GPa, 1600~\degree C, 2~h; the product was a white powder.

\subsection{Powder X-ray diffraction (PXRD)}
Powder X-ray diffraction (PXRD) was carried out on a Rigaku MiniFlex600 diffractometer with Cu K$\alpha$ radiation, $\lambda = 1.541$~\AA, range 5--70\degree, step size 0.01\degree. Rietveld refinement \cite{Rietveld1969} was carried out using \textsc{Topas} \cite{Coelho2018}. The zero offset and 12 background coefficients (Chebyshev polynomial) were refined, as were the unit cell dimensions and the $y$ fractional coordinates of the lanthanide and transition metal sites. The fractional occupancies of the transition metal site by Nb and Ta were fixed at their nominal values according to the reaction stoichiometry. The peak shape was described by a modified Thompson-Cox-Hastings pseudo-Voigt function.

\subsection{Magnetometry}
Magnetic measurements were made using a Quantum Design Materials Properties Measurement System (MPMS-3). 10--20~mg of sample (accurately measured each time) was contained in clingfilm inside a plastic straw as the sample holder. The DC magnetic moment was measured as a function of temperature upon warming in the range 1.8--300~K at a field of 500~Oe, after cooling in zero field (ZFC) or in the applied field (FC). The DC moment was also measured as a function of magnetic field in the range 0--7~T at several temperatures. The data were corrected for the presence of up to 5 wt~\%\ non-magnetic PtO (exact value determined by Rietveld refinement for each sample).

\subsection{Specific heat}
Approximately 50~mg of $M$-\ch{YbTaO4} was pressed into a pellet of diameter 5~mm and thickness about 1~mm. In order to increase the pellet's hardness, it was annealed in air for 2~h at 800~\degree C, low enough to prevent transformation to the $M'$ phase \cite{Markiv2002a} as confirmed by post-annealing PXRD. A portion of the pellet weighing 8.50~mg was attached to the sample holder using Apiezon N grease and the heat capacity was measured at $T=$ 2--60~K using a Quantum Design Physical Properties Measurement System (PPMS) EverCool II.

The lattice contribution $C_\mathrm{latt}$ was modelled using the Debye law:

\begin{equation}
C_\mathrm{latt}=\frac{9nRT^3}{{\theta_\mathrm{D}^3}}\int_{0}^{\frac{\theta_\mathrm{D}}{T}}\frac{x^4e^x}{(e^x-1)^2} dx
\end{equation}
where $n$ is the number of atoms per formula unit, $R$ is the molar gas constant, $T$ the temperature in K, and $\theta_\mathrm{D}$ the Debye temperature in K \cite{Gopal1966}. A scalar parameter $p$ was included in the fitting equation and refined to account for the sample mass error and imperfect coupling to the sample stage, since the title compound is an insulator. The lattice contribution was then subtracted from the total heat capacity to leave the magnetic contribution, $C_\mathrm{mag}(T)$.

\section{Results}

\subsection{\label{section:structure}Synthesis and structural characterisation}

Reaction of equimolar amounts of \ch{\textit{Ln}2O3} and \ch{Ta2O5} in air under ambient-pressure conditions produces $M$-\ch{\textit{Ln}TaO4} for $Ln=$ Nd--Er \cite{Kelly2022a}. However, the required synthesis temperature increases across the \textit{Ln} series, and application of pressure is necessary to favour $M$-\ch{YbTaO4} over the layered $M'$ polymorph. Brixner and Chen \cite{Brixner1983} successfully synthesised $M$-\ch{YbTaO4} at 1400~\degree C and 6~GPa, while Markiv \textit{et al.} \cite{Markiv2002a} used 1500~\degree C and 8~GPa. Our experiments found that 1~h at 6~GPa and 1800~\degree C, or alternatively 2~h at 6~GPa and 1700~\degree C, reliably produced $M$-\ch{YbTaO4} without any reflections from the competing $M'$ phase. The discrepancies between the different synthesis temperature for our study compared with the literature results is most likely related to the calibration methods of the high-pressure apparatus, where the setpoint is typically specified by an electrical power rather than an absolute temperature value. All high-pressure reactions at elevated temperatures produced small amounts of PtO from oxidation of the Pt capsule, visible as shiny grey residue on the surfaces of the sample. This proved difficult to separate from the target phase and was therefore present at up to 5 wt~\%\ in the final products. A pre-reaction at ambient pressure (to form the $M'$ phase) was also tested but was less successful than starting from the binary oxides. All synthesis attempts and results are summarised in the Supplemental Material \cite{Supplemental}.

\begin{figure}[htbp]
\centering
\includegraphics[width=0.5\textwidth]{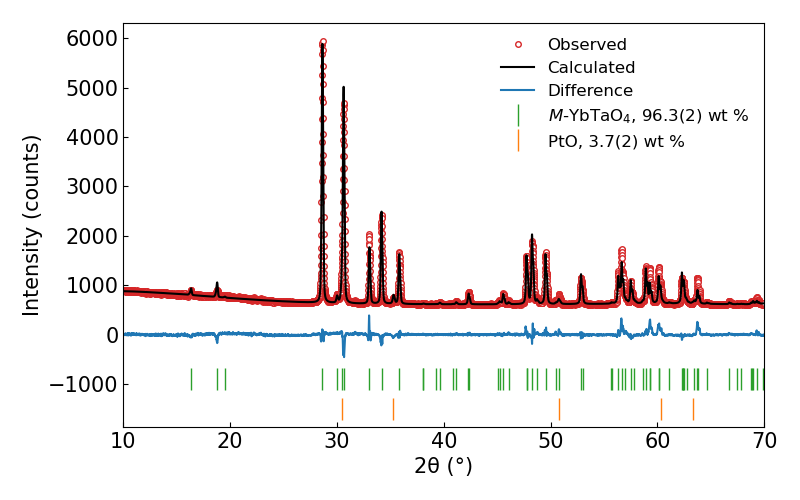}
\caption{Rietveld refinement against room-temperature PXRD data ($\lambda=$ Cu K$\alpha$, 5--70\degree, 0.01\degree\ steps) for $M$-\ch{YbTaO4}. Red circles: observed data, black line: calculated pattern, blue line: difference pattern, green tick marks: Bragg reflection positions.}
\label{fig:rietveld}
\end{figure}

\begin{figure}[htbp]
\centering
\includegraphics[width=0.5\textwidth]{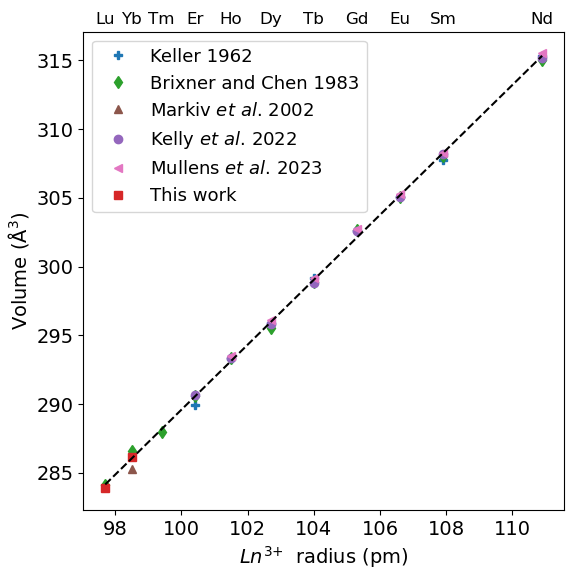}
\caption{Unit cell volume of $M$-\ch{\textit{Ln}TaO4} compounds \cite{Brixner1983,Keller1962,Kelly2022a,Markiv2002a,Mullens2023} as a function of lanthanide ionic radius \cite{Shannon1976}.}
\label{fig:volumes}
\end{figure}

The crystal structure of $M$-\ch{YbTaO4} was refined against room-temperature powder X-ray diffraction (PXRD) data using the Rietveld method \cite{Rietveld1969}. The atomic coordinates for oxide ions were fixed at values from the literature for \ch{SmTaO4} \cite{Keller1962} because of the low X-ray scattering power of O compared with Yb and Ta; the isotropic thermal parameters $U_\mathrm{iso}$ were fixed at 1~\AA$^2$ for all atoms. Table~\ref{table:ybtao4_params} gives the refined structural data for a representative sample of $M$-\ch{YbTaO4} (NK080) and Fig.~\ref{fig:rietveld} shows the refinement. We chose the $I2/a$ setting of space group 15 because it gives the conventional cell with the smallest $\beta$ angle \cite{Mighell2002}. However, other authors have previously reported the structure of $M$-\ch{YbTaO4} using the non-conventional $B2/b$ setting \cite{Markiv2002a}. After transforming that cell to $I2/a$, their reported lattice parameters agree well with our results and with those of Brixner and Chen \cite{Brixner1983}. The unit cell volume of $M$-\ch{YbTaO4} also fits into the linear trend observed for the whole $M$-\ch{\textit{Ln}TaO4} series, Fig.~\ref{fig:volumes}. We could also synthesise the non-magnetic analogue $M$-\ch{LuTaO4} using high pressure (6~GPa, 1600~\degree C, 2~h). Its structural parameters are consistent with the previous report \cite{Brixner1983} and are included in Table~\ref{table:ybtao4_params} for comparison with the Yb compound.

Refined lattice parameters for all samples of $M$-\ch{YbTaO4} from different synthesis runs are highly consistent, demonstrating that this synthesis route is repeatable and reliable (see table in the Supplemental Material \cite{Supplemental}). The crystal structure of $M$-\ch{YbTaO4} is built up from second-order Jahn-Teller distorted \ch{TaO6} octahedra \cite{Saura-Muzquiz2021}, as shown in Fig.~\ref{fig:diamond}, and \ch{YbO8} polyhedra which are best described as distorted square antiprisms with the unique direction along the $c$-axis \cite{Brixner1983}. These polyhedra are linked together by edges and vertices to form the crystal structure with its two interlocking ``stretched diamond'' networks, one for the magnetic \ch{Yb^{3+}} and one for the non-magnetic \ch{Ta^{5+}}. $M$-\ch{YbTaO4} is isostructural with \ch{YbNbO4} which has been recently investigated by Kumar \textit{et al.} \cite{Kumar2025}.

\begin{table}[htbp]
\centering
\caption{Refined structural parameters for $M$-\ch{YbTaO4} and $M$-\ch{LuTaO4} synthesised at high pressure. Space group $I2/a$, Cu~K$\alpha$ radiation, room temperature.}
\label{table:ybtao4_params}
\begin{tabular}{c c c}
\toprule
 & $M$-\ch{YbTaO4} & $M$-\ch{LuTaO4} \\
 & (NK080) & (NK060) \\
\midrule
$a$ (\AA) & 5.2692(2) & 5.25043(15) \\
$b$ (\AA) & 10.8388(4) & 10.8119(3) \\
$c$ (\AA) & 5.03159(18) & 5.02319(14) \\
$\beta$ (\degree) & 95.4562(13) & 95.3847(17) \\
$V$ (\AA$^3$)  & 286.061(19) & 283.893(14) \\
$R_\mathrm{wp}$ (\%) & 9.92 & 9.50 \\
$\chi^2$ & 2.96 & 5.94 \\
\midrule
Yb1 ($4e$) $x,y,z$ & (0.25, 0.1181(2), 0) & (0.25, 0.1189(2), 0) \\
Ta1 ($4e$) $x,y,z$ & (0.25, 0.6482(2), 0) & (0.25, 0.6460(2), 0) \\
O1 ($8f$) $x,y,z$ & (0.015, 0.719, 0.216) & (0.015, 0.719, 0.216) \\
O2 ($8f$) $x,y,z$ & (0.901, 0.456, 0.240) & (0.901, 0.456, 0.240) \\
\bottomrule
\end{tabular}
\end{table}

\begin{figure}[htbp]
\centering
\includegraphics[width=0.5\textwidth]{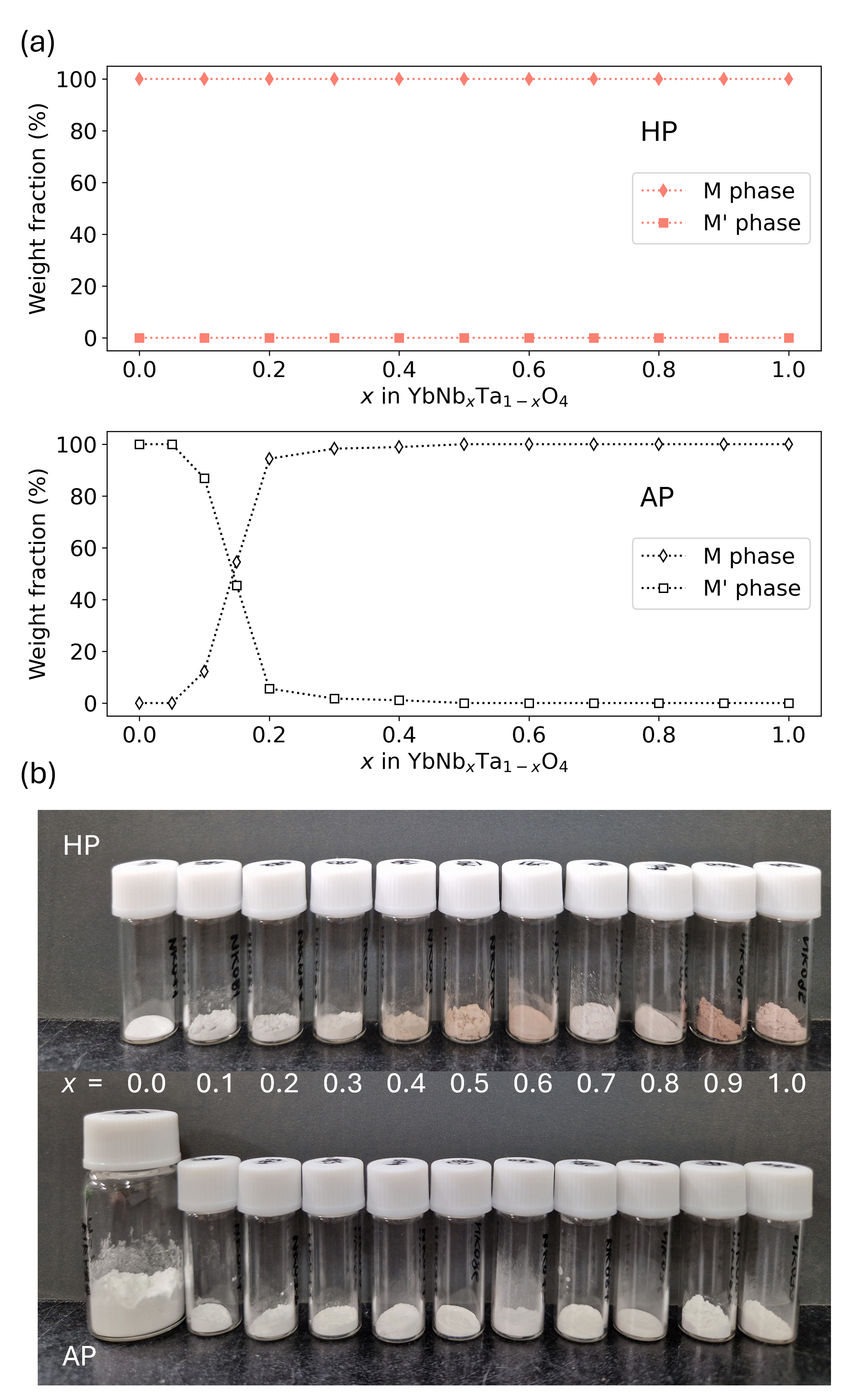}
\caption{(a) Refined weight fractions of the $M$ and $M'$ phases of \ch{YbNb_{x}Ta_{1-x}O4} made through high-pressure (HP, top) versus ambient-pressure synthesis (AP, bottom). (b) Pictures of samples show the different colours observed under different synthetic conditions.}
\label{fig:colours}
\end{figure}

\begin{figure}[htbp]
\centering
\includegraphics[width=\textwidth]{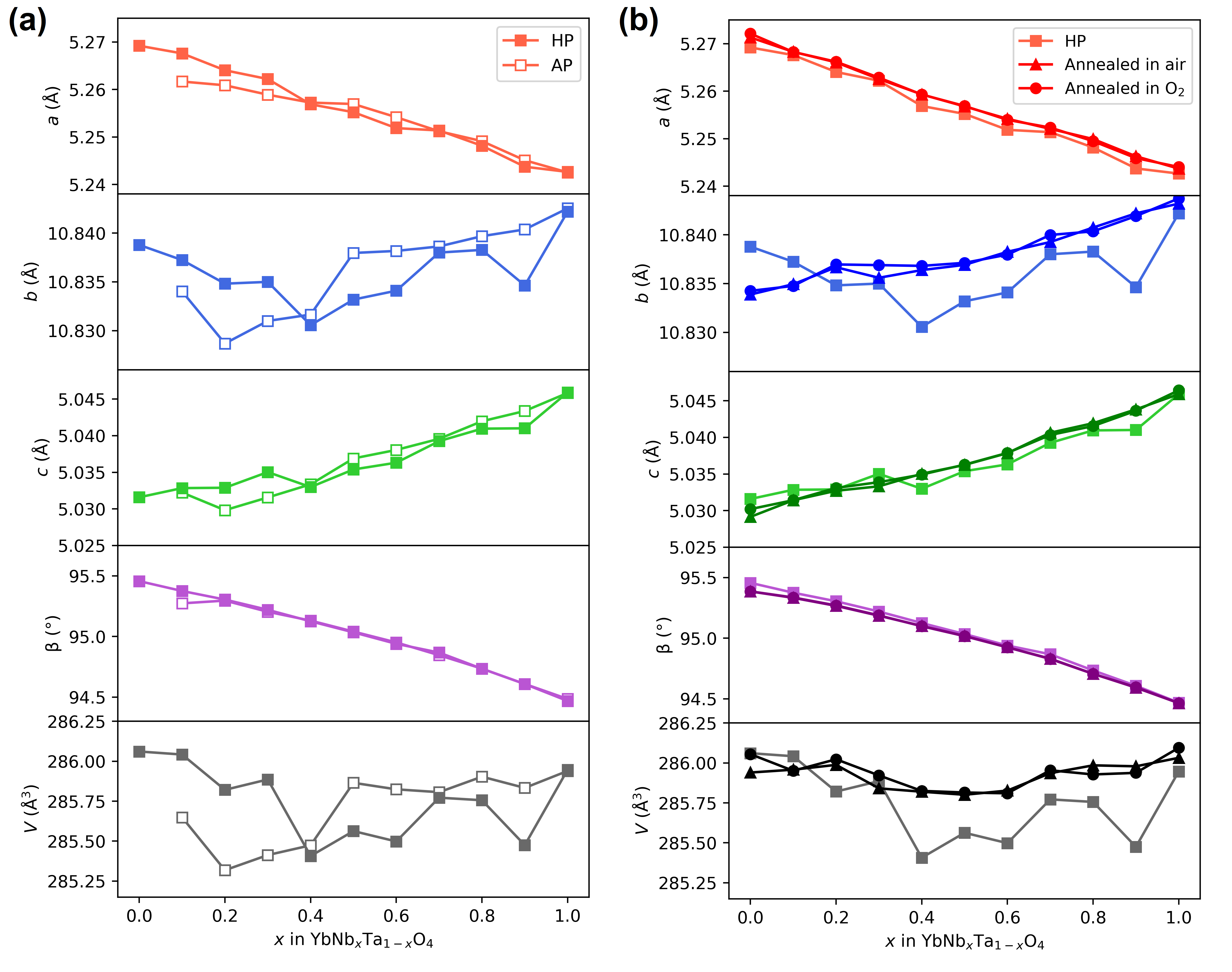}
\caption{Refined unit cell parameters of the compounds \ch{YbNb_{x}Ta_{1-x}O4} ($M$-type crystal structure) as a function of $x$. (a) Comparison of samples made by high-pressure (HP) and ambient-pressure (AP) synthesis. (b) Comparison of the HP samples as-synthesised and after annealing in either air or pure oxygen. Error bars are smaller than the datapoints in all cases.}
\label{fig:solid_solution_params}
\end{figure}

We also synthesised several samples from the solid solution \ch{YbNb_{x}Ta_{1-x}O4}, $0\leq x\leq1$. At ambient pressure, only the $M$ phase formed in the Nb-rich samples with $x\geq0.5$; only the $M'$ phase formed for the Ta-rich samples $x\leq 0.05$; while at intermediate values $0.1\leq x \leq 0.4$ a mixture of both phases was observed, Fig.~\ref{fig:colours}. Our data are consistent with the previous findings of Mullens \textit{et al.}, whose ambient-pressure synthesis at 1400~\degree C gave only the $M'$ phase for \ch{YbTaO4}, only $M$ for \ch{YbNb_{0.4}Ta_{0.6}O4} and more Nb-rich compositions, and a phase mixture for \ch{YbNb_{0.2}Ta_{0.8}O4} \cite{Mullens2023}. In contrast, under our high-pressure conditions (6~GPa, 1800~\degree C, 1~h) only the $M$ phase was observed for all $x$. However, many of the high-pressure samples with $x\neq0$ were beige in colour, rather than white like the whole ambient-pressure series (Fig.~\ref{fig:colours}). No impurities or structural transformations were visible by powder X-ray diffraction (additional Rietveld refinements available in the Supplemental Material \cite{Supplemental}) to explain the colour, which also did not seem to follow a clear trend with $x$. Furthermore, the lattice parameters were broadly consistent between the two synthesis methods, Fig.~\ref{fig:solid_solution_params}(a), within experimental error. The colour of the high-pressure samples might arise from charge transfer behaviour, i.e.~promotion of electrons from ligands (oxide ions) into the unfilled Yb $4f$ shell to form localised Yb$^{2+}$ ions \cite{Nakazawa2002}. However, we also note that off-stoichiometry in the pyrochlore oxide \ch{Yb_{$2+x$}Ti_{$2-x$}O_{$7-\delta$}} has been found to have a significant impact on the colour of single-crystal samples, even where oxygen vacancies can account for all Yb ions remaining in the $+3$ state \cite{Arpino2017}. In fact, refining the metal cation site occupancies in our samples gave no indication of off-stoichiometry. Alternatively, other defects such as nitrogen impurity centres may be present, and/or the formation of some Yb$^{2+}$ ions might be compensated by vacancies formed in the oxide-ion sublattice under high-pressure-high-temperature synthesis, which is known to alter the patterns of defects and optical centres in doped gemstones \cite{Brazhkin2007}. 

In order to investigate potential variation in the oxide stoichiometry, portions of the samples made under high pressure were subsequently annealed for 2~h at 800~\degree C in a chamber furnace under static air atmosphere or in a tube furnace under flowing \ch{O2} gas. After removal from the furnaces, all samples had changed to a pure white colour (images available in Supplemental Material \cite{Supplemental}). PXRD measurements, Fig.~\ref{fig:solid_solution_params}(b), showed that the samples annealed in air and in \ch{O2} were identical to each other and their lattice parameters were very close to the as-made samples, indicating that there was no structural transition. Overall, the trends with $x$, particularly the trend in $b$, became smoother after annealing. The low scattering power of oxygen meant that the oxygen site occupancies could not be refined. However, these observations suggest that the observed colour change upon annealing is related to a change in the oxide ion composition and/or the oxidation of isolated Yb$^{2+}$ ions. The presence of Yb$^{2+}$ ions in the as-made samples would be possible because of the slightly reducing synthetic environment of the high-pressure apparatus (Pt capsule surrounded by graphite).

\subsection{\label{section:magnetic}Magnetic measurements}

Magnetic susceptibility measurements were carried out on several samples of $M$-\ch{YbTaO4} (as-made after HP synthesis). The presence of nonmagnetic PtO up to about 5 wt \%, as mentioned in Section~\ref{section:structure}, was quantified by Rietveld refinement \cite{Supplemental} in order to correct the raw data. The susceptibility is given by $\chi\approx M/H$ because $H$ is small (500~Oe) and $M(H)$ is linear in this region. No magnetic ordering was observed in any of the samples down to a minimum temperature of 1.8~K and the ZFC and FC curves were identical (Supplemental Material \cite{Supplemental}), which rules out any spin-glass behaviour in this temperature range. A representative dataset is shown in Fig.~\ref{fig:mag_combined}(a) and fitted parameters for all samples are tabulated in the Supplemental Material \cite{Supplemental}.

\begin{figure}[htbp]
\centering
\includegraphics[width=\textwidth]{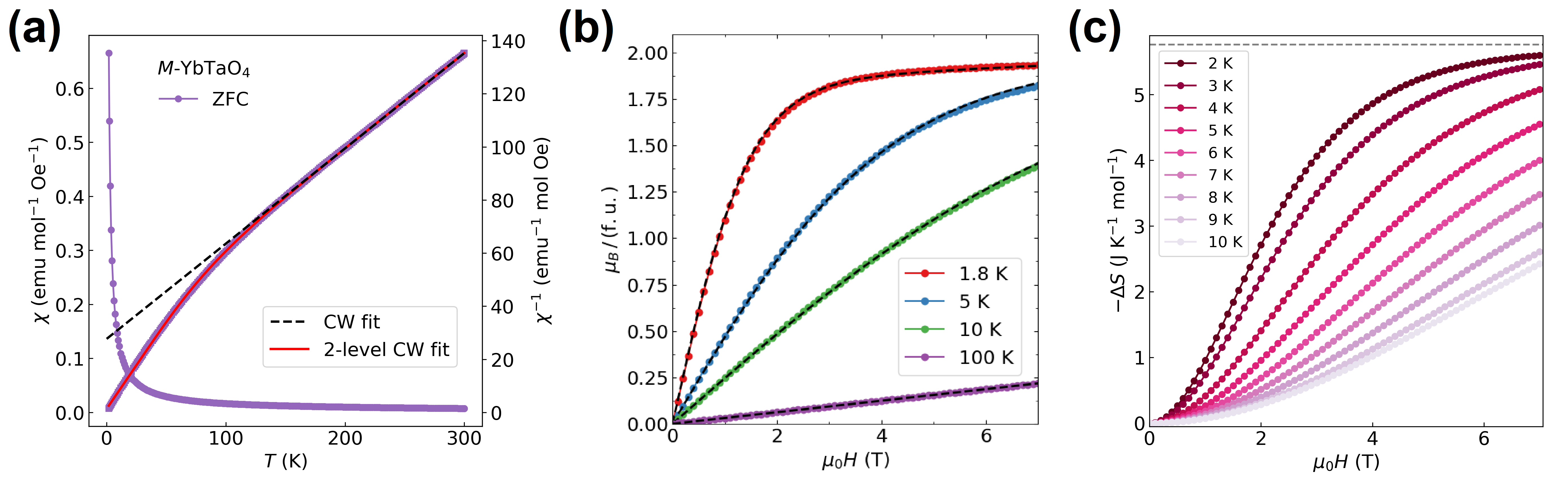}
\caption{Magnetic data for a sample of $M$-\ch{YbTaO4} (NK099): (a) Magnetic susceptibility $\chi(T)$ and inverse susceptibility $\chi^{-1}(T)$. The inverse susceptibility was fitted to a simple Curie-Weiss law (black dashed line, extrapolated from 150 down to 0~K) or a modified two-level Curie-Weiss law to account for crystal electric field effects (red solid line). (b) Isothermal magnetisation curves, fitted to the Brillouin function with $J=\frac{1}{2}$ (black dashed lines). (c) Magnetocaloric effect calculated from magnetic isotherms. The grey dashed line shows the expected maximum value of $-\Delta S=R\ln2$.}
\label{fig:mag_combined}
\end{figure}

The inverse susceptibility, $\chi^{-1}(T)$, was first fitted to the Curie-Weiss (CW) law:

\begin{equation}
\frac{1}{\chi}=\frac{8(T-\theta_\mathrm{CW})}{\mu_\mathrm{eff}^2}
\end{equation}
where $\theta_\mathrm{CW}$ is the Weiss temperature and $\mu_\mathrm{eff}$ the effective magnetic moment. Linear fitting in the range 150--300~K, on average for several samples of $M$-\ch{YbTaO4}, yielded $\theta_\mathrm{CW}=-78(7)$~K and $\mu_\mathrm{eff}=4.79(11)$~$\mu_\mathrm{B}$ (black dashed line in Fig.~\ref{fig:mag_combined}(a)). Whilst the effective moment is close to the estimated free-ion value for \ch{Yb^{3+}} ions, 4.54~$\mu_\mathrm{B}$, the Weiss temperature is unphysically large for the expected strength of magnetic exchange interactions, which is typically of the order of 1--10~K for lanthanide compounds. Similarly large estimates of the Weiss temperature have been observed for other Yb compounds and the unsuitability of the simple Curie-Weiss model is explained by contributions from thermally populated crystal electric field (CEF) levels above the $J=\frac{7}{2}$ ground state; a low-temperature Curie-Weiss fit is often used to mitigate this \cite{Pan2024,Jiang2022}. However, Yb$^{3+}$ is a Kramers ion (with an odd number of unpaired electrons) so its $J=\frac{7}{2}$ state is expected to split into four doublets, with the ground state being a $J=\pm\frac{1}{2}$ doublet, separated from the first excited state (a $J=\pm\frac{3}{2}$ doublet) by an energy gap $\Delta E_{10}$. Therefore, we fitted the data in the entire temperature range (1.8--300~K) to a two-level function:

\begin{equation}
\frac{1}{\chi}=8(T-\theta_\mathrm{CW})\left(\frac{1+e^{-\Delta E_{10}/k_\mathrm{B}T}}{{\mu_0}^2 + {\mu_1}^2e^{-\Delta E_{10}/k_\mathrm{B}T}}\right)
\end{equation}
where $\mu_0$ is the effective magnetic moment in the ground state, $\mu_1$ the effective moment in the first excited state, and $\Delta E_{10}$ the energy difference between those states \cite{Mugiraneza2022}. The two-level fitting (red solid line in Fig.~\ref{fig:mag_combined}(a)) was carried out for eleven distinct $M$-\ch{YbTaO4} samples from different synthesis runs (tabulated in the Supplemental Material \cite{Supplemental}). The fitted parameters, on average over all the samples, were $\theta_\mathrm{CW}=-2.4(6)$~K, $\Delta E_{10}=256(16)$~K, $\mu_0=3.48(8)\ \mu_\mathrm{B}$ and $\mu_1=5.8(3)\ \mu_\mathrm{B}$. The negative Weiss temperature indicates antiferromagnetic interactions in $M$-\ch{YbTaO4} and the sizes of the effective moments are in line with those obtained in other Yb-based metal oxides using the same fitting model \cite{Kumar2024,Kumar2025,Kumar2025a}. We also fitted the susceptibility to the CW law in the low-temperature regime, $1.8\leq T(\mathrm{K})\leq 30$, which resulted in $\theta_\mathrm{CW}^\mathrm{LT}=-0.68(14)$~K and $\mu_\mathrm{eff}^\mathrm{LT}=3.32(8)\ \mu_\mathrm{B}$. The obtained low-temperature effective moment is clearly reduced compared with the high-temperature effective moment, and importantly it is consistent with $\mu_0$ obtained from the two-level CW fitting, indicating that the two-level $J_\mathrm{eff}=\frac{1}{2}$ model is a good fit for our data.

Isothermal magnetisation data collected on $M$-\ch{YbTaO4} as a function of field are shown in Fig.~\ref{fig:mag_combined}(b). The magnetisation approached saturation at low temperatures and high fields. The data at 1.8~K were first fitted to a linear function in the range 5--7~T, representing the temperature-independent paramagnetism \cite{Guchhait2025}. This yielded a gradient $\chi_\mathrm{0}=1.5(2)\times10^{-2}$~emu~mol$^{-1}$ and intercept (saturation magnetisation) $\mu_\mathrm{sat}=1.82(6)$~$\mu_\mathrm{B}$. The $g$-factor could then be estimated using $g_\mathrm{eff}=(\mu_\mathrm{sat}/\mu_\mathrm{B})/J_\mathrm{eff}$ \cite{Arjun2023} $=3.65(12)$ for $J_\mathrm{eff}=\frac{1}{2}$.

Next, the datasets at each temperature $T$ were fitted in the entire field range to the function:
\begin{equation}
M(H) = \chi_\mathrm{0}H + gJB_J(H)
\end{equation}
where $g$ is the Land\'{e} g-factor and $B_J(H)$ the Brillouin function. For an effective $J=\frac{1}{2}$, typical of \ch{Yb^{3+}} compounds, the Brillouin function is given by $B_J(H)=\tanh{(\frac{gJ\mu_\mathrm{B}H}{k_\mathrm{B}T})}$. The fits are shown in Fig.~\ref{fig:mag_combined}(b) as black dashed lines. The fitted $g$-factor at 1.8~K was $g=3.58(8)$, which is consistent with our earlier estimate and comparable with those of \ch{NaYbGeO4} (3.84) \cite{Arjun2023}, $M'$-\ch{YbTaO4} (3.2) \cite{Kumar2024}, \ch{YbNbO4} (3.0) \cite{Kumar2025}, and \ch{Na_{0.5}Yb_{0.5}WO4} (3.1) \cite{Kumar2025a}. The saturation magnetisation $M_\mathrm{sat}$, estimated by extrapolating the high-field linear region of $M(H)$ back to zero field, was 1.82(6)~$\mu_\mathrm{B}$. Comparing this value with $g_J.J=4$ for the Yb$^{3+}$ ion indicates substantial single-ion anisotropy, most likely Ising-like because $M_\mathrm{sat}\approx g_J.J/2$ \cite{Bramwell2000}.

We also calculated the magnetocaloric effect (MCE) from evenly spaced magnetic isotherms (with datapoints at 1~K intervals from 2--10~K and 1000~Oe intervals from 0--7~T) using Maxwell's relations \cite{Pecharsky1999}. We found a maximum MCE of $-\Delta S_\mathrm{mag}(2\mathrm{K},7\mathrm{T})=5.60$~J~mol$^{-1}$~K$^{-1}$, which is almost equal to $R\ln2=5.76$~J~mol$^{-1}$~K$^{-1}$, Fig.~\ref{fig:mag_combined}(c). This indicates that the effective quantum number $J_\mathrm{eff}=\frac{1}{2}$ is appropriate for $M$-\ch{YbTaO4}, as for a number of other Yb-based oxides \cite{Ramanathan2024,Kumar2024,Kumar2025,Kumar2025a,Arjun2023}.

Magnetic susceptibility and isothermal magnetisation data were also collected on all samples from the solid solution \ch{YbNb_{x}Ta_{1-x}O4} after annealing in oxygen (see Section \ref{section:structure}) and on several other samples either as-made or annealed in air. Two-level Curie-Weiss fits to the susceptibility were carried out in the same manner as described above. The fitted magnetic parameters (Supplemental Material \cite{Supplemental}) showed no obvious trends as a function of $x$ and the spread of data for each parameter was of a similar magnitude to the uncertainties reported above, which were obtained by averaging over several samples of the same composition. Therefore, we cannot draw any conclusions from this dataset about the effect of Nb substitution on the magnetic properties of $M$-\ch{YbTaO4}.

\subsection{\label{section:hc}Specific heat}
\begin{figure}[htbp]
\centering
\includegraphics[width=\textwidth]{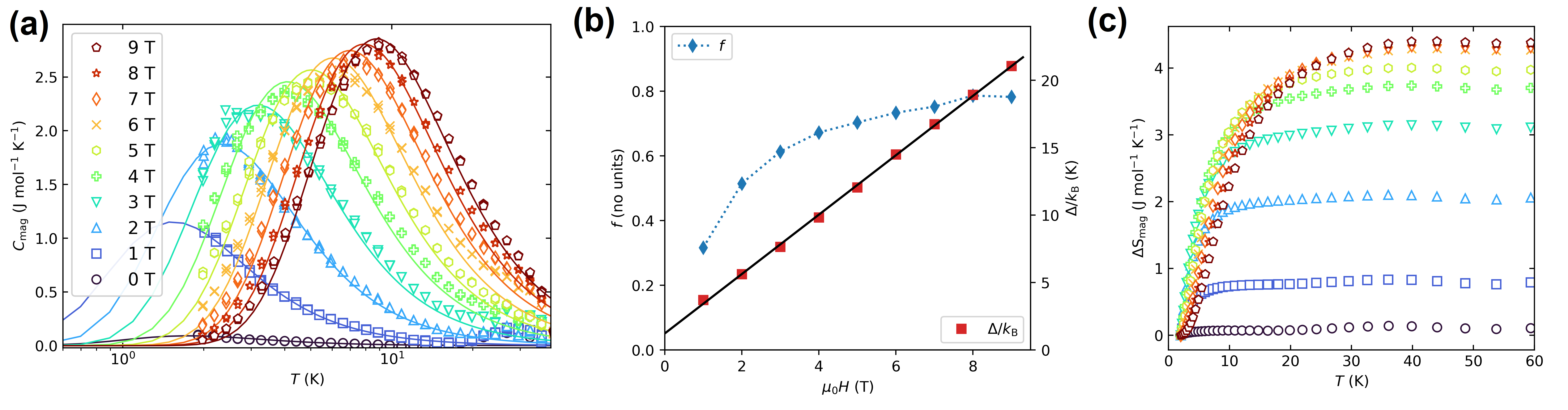}
\caption{(a) Magnetic specific heat $C_\mathrm{mag}$ for $M$-\ch{YbTaO4} at different applied fields, fitted to the two-level Schottky equation. (b) Fitted parameters $f$ and $\Delta/k_\mathrm{B}$ as a function of field. (c) Magnetic entropy change obtained by integration of $C_\mathrm{mag}/T$; for the legend, see part (a).}
\label{fig:specific_heat}
\end{figure}

Specific heat data were collected on $M$-\ch{YbTaO4} at several fields in the temperature range 2--60~K and fitted to the Debye law with $\theta_\mathrm{D}=246(2)$~K in order to provide an estimate of the lattice contribution to the specific heat, $C_\mathrm{latt}$, which was then subtracted to give the magnetic contribution $C_\mathrm{mag}$. Fig.~\ref{fig:specific_heat}(a) shows $C_\mathrm{mag}$ for all fields. In non-zero applied fields a broad Schottky-type anomaly is visible, shifting to higher temperatures as the field increases. This behaviour is consistent with Zeeman splitting of a $J=\frac{1}{2}$ doublet \cite{Guchhait2025}. The $C_\mathrm{mag}$ curves were fitted to a two-level Schottky function:
\begin{equation}
C_\mathrm{S}(T,H)=fR\left(\frac{\Delta}{k_\mathrm{B}T}\right)^2\frac{e^{\Delta/k_\mathrm{B}T}}{(e^{\Delta/k_\mathrm{B}T}+1)^2}
\end{equation}
where $f$ is the molar fraction of free spins, $R$ the molar gas constant, and $\Delta$ the Zeeman energy gap in the split ground state doublet. The model fits our data well, as shown by the solid lines in Fig.~\ref{fig:specific_heat}(a). The gap $\Delta$ increases linearly with field as expected, with $f$ also increasing but not reaching 100~\%\ by $\mu_0H=9$~T, Fig.~\ref{fig:specific_heat}(b). Extrapolation of the straight-line fit of $\Delta(H)$ indicates a small finite energy gap at zero field, $\Delta(0)\approx1.2$~K, which may correspond to interactions between Yb$^{3+}$ ions, but is well below the minimum temperature of the present work. Further measurements at $T<2$~K will be important for future investigation of $M$-\ch{YbTaO4}.

The magnetic entropy, $\Delta S_\mathrm{mag}$, was obtained by integrating $C_\mathrm{mag}/T$ and is shown in Fig.~\ref{fig:specific_heat}(c). At high fields the entropy saturates at 4.38~J~mol$^{-1}$~K$^{-1}$, which is around three-quarters of the expected value of $R\ln2=5.76$ J~mol$^{-1}$~K$^{-1}$ for a two-level system. Since we found that the magnetic entropy change reached almost $R\ln2$ in our MCE calculation (Fig.~\ref{fig:mag_combined}(c)), this is probably a systematic offset with the ``missing'' entropy being gained below 2~K, because the starting point for our numerical integration of $C_\mathrm{mag}/T$ was $T=2$~K instead of the ideal value of zero.

\section{\label{section:discussion}Discussion}
The lanthanide niobates and tantalates exhibit several different structural polymorphs including at least one tetragonal and three monoclinic phases, depending on the transition metal ion (Nb vs Ta), lanthanide ion, temperature, and pressure \cite{Mullens2023}. Of these, the $M$ or fergusonite-type monoclinic phase is of particular interest for its stretched diamond network of magnetic ions, which has previously been studied in both the niobates \cite{Kumar2025} and tantalates \cite{Kelly2022a,Zhang2025}. While antiferromagnetically coupled spins on a cubic diamond lattice are expected to display long-range magnetic order, lowering of symmetry to tetragonal or monoclinic crystal structures typically introduces magnetic frustration due to competing interactions. Additional complexity is possible in the case of non-magnetic substitution onto the diamond-like lattice, for example in \ch{Na_{0.5}Yb_{0.5}WO4} with 50:50 random site occupancy of Na$^+$ and Yb$^{3+}$ \cite{Kumar2025a}.

$M$-\ch{YbTaO4} is an example of a monoclinic stretched diamond lattice with no site disorder. Like its niobate analogue \ch{YbNbO4} \cite{Kumar2025}, we observed evidence for a $J_\mathrm{eff}=\frac{1}{2}$ Kramers doublet down to 1.8~K in the analysis of susceptibility, isothermal magnetisation, and specific heat data. Kramers behaviour is typical of compounds containing the Yb$^{3+}$ ion, which has a half-integer spin and strong spin-orbit coupling, in a low-symmetry crystal electric field environment such as the distorted \ch{YbO8} polyhedra of $M$-\ch{YbTaO4}. In such compounds, the $J_\mathrm{eff}=\frac{1}{2}$ ground state means that the effective magnetic moment at low temperatures is smaller than expected. Indeed, the susceptibility data for $M$-\ch{YbTaO4} clearly showed that the Curie-Weiss law does not hold over the entire temperature range studied (1.8--300~K), i.e., the effective magnetic moment is strongly temperature-dependent. We calculated $\theta_\mathrm{CW}^\mathrm{LT}=-0.68(14)$~K, which suggests that the exchange interactions between neighbouring Yb$^{3+}$ ions are very weak. The nearest-neighbour interaction strength $J_\mathrm{nn}$, in the mean-field limit, may be estimated as follows \cite{Ramirez1994}:
\begin{equation}
J_\mathrm{nn}\approx\frac{3k_\mathrm{B}\theta_\mathrm{CW}}{2zS(S+1)}
\end{equation}

Substituting $J_\mathrm{eff}=\frac{1}{2}$ for the spin quantum number $S$, and setting the number of neighbouring ions $z=4$, we obtain an estimated $J_\mathrm{nn}\approx-0.34$~K using $\theta_\mathrm{CW}^\mathrm{LT}$. We can also estimate the dipolar interaction, $D=-\mu_0\mu_\mathrm{eff}^2/4\pi r^3$, where $r$ is the (average) nearest-neighbour distance and $\mu_0=4\pi\times10^{-7}$ is the permeability of free space \cite{Gingras2011}. Each Yb$^{3+}$ ion in $M$-\ch{YbTaO4} has four neighbouring Yb$^{3+}$ ions, two at a distance of 3.691(4)~\AA\ and two at a distance of 3.791(4)~\AA\ according to our Rietveld refinements (at room temperature). Taking the average as $r\approx3.741$~\AA, we calculate $D\approx-0.29$~K which is a similar size to $J_\mathrm{nn}$. However, these mean-field approximations do not take spin anisotropy into account, whereas our $M(H)$ data indicate substantial anisotropy for the Yb$^{3+}$ ions. Both of the estimated values are considerably smaller than 1.8~K, so the effects of the interplay between dipolar and exchange coupling will only be observed at temperatures below the minimum achieved in this experimental study. 

Magnetocaloric calculations for $M$-\ch{YbTaO4} indicate that the magnetic entropy change is more than 97~\%\ of the maximum ($R\ln2$) at 2~K and 7~T. This result, coupled with the weak magnetic exchange and geometric frustration which should act to suppress magnetic ordering, means that $M$-\ch{YbTaO4} has the potential to be used as a magnetocaloric material in adiabatic demagnetisation refrigerators (ADRs) at liquid-helium or dilution temperatures. ADRs have traditionally used single crystals of paramagnetic salts, but these often contain water of crystallisation and therefore cannot be evacuated or heated above 100~\degree C in preparation for ultrahigh vacuum applications. However, recent studies have found that some chemically stable ytterbium-based oxides are strong contenders for new ADR technology \cite{Tokiwa2021,Arjun2023,Arjun2023a}. The ADR technique itself can even be used to determine magnetic transition temperatures, using an indirect measurement of heat capacity calculated from the heat input and the time derivative of the measured temperature \cite{Arjun2023}. The experimental setup for such a measurement usually requires at least 2 grams of the sample, which would necessitate the combination of several 0.5~g samples from different high-pressure synthesis runs. However, our results show that this synthesis route is repeatable and produces samples with consistent structural and magnetic parameters, so there is potential for $M$-\ch{YbTaO4} to be investigated for ADR applications in the near future.

Finally, we compare $M$-\ch{YbTaO4} with the niobate \ch{YbNbO4}. The two compounds have almost identical unit cell dimensions, less than 0.05~\%\ difference in unit cell volume, because Nb$^{5+}$ and Ta$^{5+}$ have identical ionic radii (0.64~\AA\ if 6-coordinate \cite{Shannon1976}). Given the very similar ionic radii, it is perhaps surprising that high temperature and/or pressure (depending on \textit{Ln}) is required to stabilise $M$ over $M'$ for the lanthanide tantalates whilst the niobates do not display the $M'$ phase at all. The \ch{HoTaO4}-\ch{HoNbO4} solid solution has recently been explored in depth using a combined synchrotron X-ray and neutron diffraction study \cite{Mullens2023}. The authors found subtle differences in the (Nb/Ta)--O bond lengths which could be related to the natures of the valence orbitals ($4d$ vs $5d$), with Ta--O bonds having greater covalency than Nb--O bonds. The transition metal ions were also more displaced from the centres of the \ch{(Nb/Ta)O6} polyhedra in the Ta-rich samples, which was described as a second-order Jahn-Teller effect. Since the phase behaviour of \ch{\textit{Ln}(Nb/Ta)O4} compounds depends strongly on the \textit{Ln} ion \cite{Brixner1983,Markiv2002a}, and oxide ions are weak scatterers of X-rays, we plan to carry out neutron diffraction studies on our Yb samples in order to locate the oxide ions accurately and explore this further.

\ch{YbNbO4}, recently investigated by Kumar \textit{et al.}, also has a $J_\mathrm{eff}=\frac{1}{2}$ Kramers doublet ground state with temperature-dependent effective magnetic moment. In \ch{YbNbO4}, muon-spin relaxation ($\mu$SR) data at low temperatures demonstrated strong magnetic correlations in a dynamic internal field with no ordering or freezing of spins down to 300~mK, making it a strong candidate for quantum spin liquid properties \cite{Kumar2025}. Our results at $T\geq1.8$~K suggest that $M$-\ch{YbTaO4} would also benefit from $\mu$SR and/or neutron diffraction measurements at low temperatures to rule out long- or short-range magnetic order, and inelastic scattering at higher temperatures to investigate the crystal electric field levels and confirm the size of the energy gap between the ground and excited state doublets. Further specific heat and AC susceptibility experiments below 1.8~K are also planned.

\section{\label{section:conclusion}Conclusions}
We synthesised $M$-\ch{YbTaO4}, which can only be produced under high-pressure-high-temperature conditions, and measured its bulk magnetic properties. Similarly to its niobate analogue, which also hosts a stretched diamond lattice of Yb$^{3+}$ spins, no magnetic ordering or spin freezing is observed down to a minimum temperature of 1.8~K. Analysis of specific heat data provides evidence for a $J_\mathrm{eff}=\frac{1}{2}$ Kramers doublet state down to 1.8~K, which splits upon application of an external magnetic field. This conclusion is further supported by analysis of the low-temperature magnetic susceptibility, magnetic isotherms and magnetocaloric calculations. 

In the solid solution \ch{YbNb_{x}Ta_{1-x}O4}, the application of pressure is required at $x\leq0.1$ to stabilise the $M$ phase at all, and at $x\leq0.4$ to obtain only the $M$ phase. Under conditions of 1800~\degree C and 6~GPa, the competing $M'$ phase is completely suppressed for all $x$. However, an interesting variation in colour across the series of high-pressure samples was observed and ascribed to off-stoichiometry, pointing to subtle differences in the crystal chemistry which should be investigated further using high-resolution diffraction techniques and optical measurements.

Overall, this study has explored a new material example of spin-$\frac{1}{2}$ magnetism on a diamond-like lattice, synthesised in a practical high-pressure synthesis window with high reproducibility. $M$-\ch{YbTaO4} has possible future applications in ADR technology, owing to its chemical stability and lack of magnetic ordering above 1.8~K, and has a high potential for unusual quantum magnetism to be observed at low temperatures in future studies.       

\begin{acknowledgments}
We acknowledge funding from Jesus College, Cambridge for a Research Fellowship, from the EPSRC for the use of the Advanced Materials Characterisation Suite (EP/M000524/1), and from the Japan Society for the Promotion of Science (JSPS) under the Summer Program international research fellowship (SP25116) and KAKENHI Grants-in-Aid for Scientific Research (25K01507 and 25K01657). Part of this work was supported by ``Advanced Research Infrastructure for Materials and Nanotechnology in Japan (ARIM)'' of the Ministry of Education, Culture, Sports, Science and Technology (MEXT), proposal number JPMXP1225NM5105.

\textbf{CRediT Author Contributions:} NDK: Conceptualization, Funding acquisition, Investigation, Formal analysis, Visualization, Writing - original draft. XL: Investigation, Writing - review and editing. SED: Conceptualization, Resources, Supervision, Writing - review and editing. KY: Funding acquisition, Resources, Writing - review and editing. YT: Funding acquisition, Resources, Supervision, Writing - review and editing.
\end{acknowledgments}

\bibliography{library}

\appendix

\renewcommand{\thefigure}{S\arabic{figure}}
\setcounter{figure}{0}
\renewcommand{\thetable}{S\arabic{table}}
\setcounter{table}{0}

\begin{table}[htbp]
\centering
\caption{All synthesis attempts to form \ch{YbTaO4}. All samples were prepared from the constituent oxides (\ch{Yb2O3} and \ch{Ta2O5}), \textit{except} the samples marked with a dagger(\textdagger), which were prepared from $M'$-\ch{YbTaO4} precursor (NK067) pre-synthesised in a furnace at ambient pressure.}
\label{table:synthesis_attempts_ybtao}
\begin{tabular}{l | c c c | c c c}
\toprule
Sample & $P$ (GPa) & $T$ (\degree C) & Time (h) & $M$ (\%) & $M'$ (\%) & Other \\
\midrule
NK067 & Ambient & 1500 & 24 & 0 & 100 & -- \\
NK068 & Ambient & 1500 & 24 & 0 & 100 & -- \\
\midrule
NK053 & 6 & 1200 & 4 & 0 & 0 & \ch{Yb3TaO7} + \ch{Ta2O5} \\
NK058 & 6 & 1400 & 1 & 2 & 98 & -- \\
NK063 & 6 & 1500 & 1 & 5 & 95 & -- \\
NK059 & 6 & 1600 & 2 & 50 & 50 & -- \\
NK064 & 6 & 1700 & 1 & 97 & 1 & 2~\%\ \ch{Ta2O5} \\
NK070(\textdagger) & 6 & 1700 & 1 & 85 & 14 & 1~\%\ \ch{Yb2O3} \\
NK071 & 6 & 1700 & 2 & 95.1 & 0 & 4.9~\%\ PtO \\
NK072(\textdagger)  & 6 & 1700 & 2 & 96.0 & 3.0 & 1.0~\%\ \ch{Yb2O3} \\
\midrule
NK078  & 6 & 1800 & 1 & 96.3 & 0 & 3.7~\%\ PtO \\
NK079  & 6 & 1800 & 1 & 97.1 & 0 & 2.9~\%\ PtO \\
NK080  & 6 & 1800 & 1 & 95.1 & 0 & 4.9~\%\ PtO \\
NK096  & 6 & 1800 & 1 & 94.8 & 0 & 3.6~\%\ PtO, 1.6~\%\ \ch{Ta2O5} \\
NK097  & 6 & 1800 & 1 & 96.7 & 0 & 2.8~\%\ PtO, 0.5~\%\ \ch{Ta2O5} \\
NK098A & 6 & 1800 & 1 & 98.0 & 0 & 2.0~\%\ PtO \\
NK099  & 6 & 1800 & 1 & 97.0 & 0 & 3.0~\%\ PtO \\
NK100  & 6 & 1800 & 1 & 96.0 & 0 & 4.0~\%\ PtO \\
NK101  & 6 & 1800 & 1 & 97.7 & 0 & 2.3~\%\ PtO \\
NK102  & 6 & 1800 & 1 & 98.4 & 0 & 1.6~\%\ PtO \\
\bottomrule
\end{tabular}
\end{table}

\begin{table}[htbp]
\centering
\caption{All synthesis attempts to form Nb-doped samples \ch{YbTa_{1-x}Nb_{x}O4}. All samples were prepared from the constituent oxides \ch{Yb2O3}, \ch{Ta2O5} and \ch{Nb2O5}.}
\label{table:synthesis_attempts_nbdoped}
\begin{tabular}{l | c | c c c | c c c}
\toprule
Sample & Target $x$ & $P$ (GPa) & $T$ (\degree C) & Time (h) & $M$ (\%) & $M'$ (\%) & Other \\
\midrule
NK081 &  0.1 & 6 & 1800 & 1 & 97.6 & 0 & 2.4~\%\ PtO \\
NK082 &  0.2 & 6 & 1800 & 1 & 97.8 & 0 & 2.2~\%\ PtO \\
NK083 &  0.3 & 6 & 1800 & 1 & 99.2 & 0 & 0.8~\%\ PtO \\
NK089 &  0.4 & 6 & 1800 & 1 & 99.4 & 0 & 0.6~\%\ PtO \\
NK090 &  0.5 & 6 & 1800 & 1 & 100 & 0 & -- \\
NK091 &  0.6 & 6 & 1800 & 1 & 100 & 0 & -- \\
NK092 &  0.7 & 6 & 1800 & 1 & 100 & 0 & -- \\
NK093A & 0.8 & 6 & 1800 & 1 & 100 & 0 & -- \\
NK094 &  0.9 & 6 & 1800 & 1 & 100 & 0 & -- \\
NK095 &  1.0 & 6 & 1800 & 1 & 100 & 0 & -- \\
\midrule
NK067 & 0    & Ambient & 1500 & 24 & 0     & 100 & -- \\
NK103 & 0.05 & Ambient & 1500 & 24 & 0     & 100 & -- \\
NK084 & 0.1  & Ambient & 1500 & 24 & 12.2  & 86.8 & -- \\
NK104 & 0.15 & Ambient & 1500 & 24 & 54.5  & 45.5 & -- \\
NK073 & 0.2  & Ambient & 1500 & 24 & 94.4  & 5.6 & -- \\
NK085 & 0.3  & Ambient & 1500 & 24 & 98.3  & 1.7 & -- \\
NK074 & 0.4  & Ambient & 1500 & 24 & 98.9  & 1.1 & -- \\
NK086 & 0.5  & Ambient & 1500 & 24 & 100   & 0 & -- \\
NK075 & 0.6  & Ambient & 1500 & 24 & 100   & 0 & -- \\
NK087 & 0.7  & Ambient & 1500 & 24 & 100   & 0 & -- \\
NK076 & 0.8  & Ambient & 1500 & 24 & 100   & 0 & -- \\
NK088 & 0.9  & Ambient & 1500 & 24 & 100   & 0 & -- \\
NK077 & 1.0  & Ambient & 1500 & 24 & 100   & 0 & -- \\
\bottomrule
\end{tabular}
\end{table}

\begin{table}[htbp]
\centering
\caption{Refined unit cell parameters for $M$-\ch{YbTaO4} samples synthesised at high pressure. Space group $I2/a$, Cu~K$\alpha$ radiation, room temperature.}
\label{table:ybtao4_params_allsamples}
\begin{tabular}{l | c c c c c c c}
\toprule
Sample & $a$ (\AA) & $b$ (\AA) & $c$ (\AA) & $\beta$ (\degree) & $V$ (\AA$^3$) & $R_\mathrm{wp}$ (\%) & $\chi^2$ \\
\midrule
NK071 & 5.26877(11) & 10.8351(2) & 5.03043(9) & 95.4500(10) & 285.877(9) & 9.80 & 2.81 \\
NK078 & 5.26797(15) & 10.8371(3) & 5.03117(13) & 95.4400(13) & 285.933(13) & 10.13 & 2.72 \\
NK079 & 5.26751(14) & 10.8324(3) & 5.02981(12) & 95.4365(12) & 285.709(12) & 9.66 & 2.88 \\
NK080 & 5.2692(2) & 10.8388(4) & 5.03159(18) & 95.4562(13) & 286.061(19) & 9.92 & 2.96 \\
NK096 & 5.27124(11) & 10.8402(2) & 5.03350(9) & 95.4446(10) & 286.322(10) & 9.96 & 2.99 \\
NK097 & 5.27061(10) & 10.83847(19) & 5.03253(8) & 95.4446(9) & 286.188(9) & 9.71 & 2.92 \\
NK098A & 5.26978(13) & 10.8379(3) & 5.03232(11) & 95.4480(11) & 286.115(12) & 10.77 & 3.56 \\
NK099 & 5.27048(13) & 10.8378(3) & 5.03233(11) & 95.4462(12) & 286.152(12) & 10.85 & 3.63 \\
NK100 & 5.26896(15) & 10.8382(3) & 5.03216(12) & 95.4534(14) & 286.066(13) & 10.41 & 2.92 \\
NK101 & 5.27149(10) & 10.83862(17) & 5.03313(8) & 95.4507(11) & 286.271(8) & 11.49 & 4.09 \\
NK102 & 5.27046(9) & 10.8371(2) & 5.03250(9) & 95.4466(9) & 286.140(9) & 10.14 & 2.89 \\
\bottomrule
\end{tabular}
\end{table}

\newpage

\begin{figure}[htbp]
\centering
\includegraphics[width=0.75\textwidth]{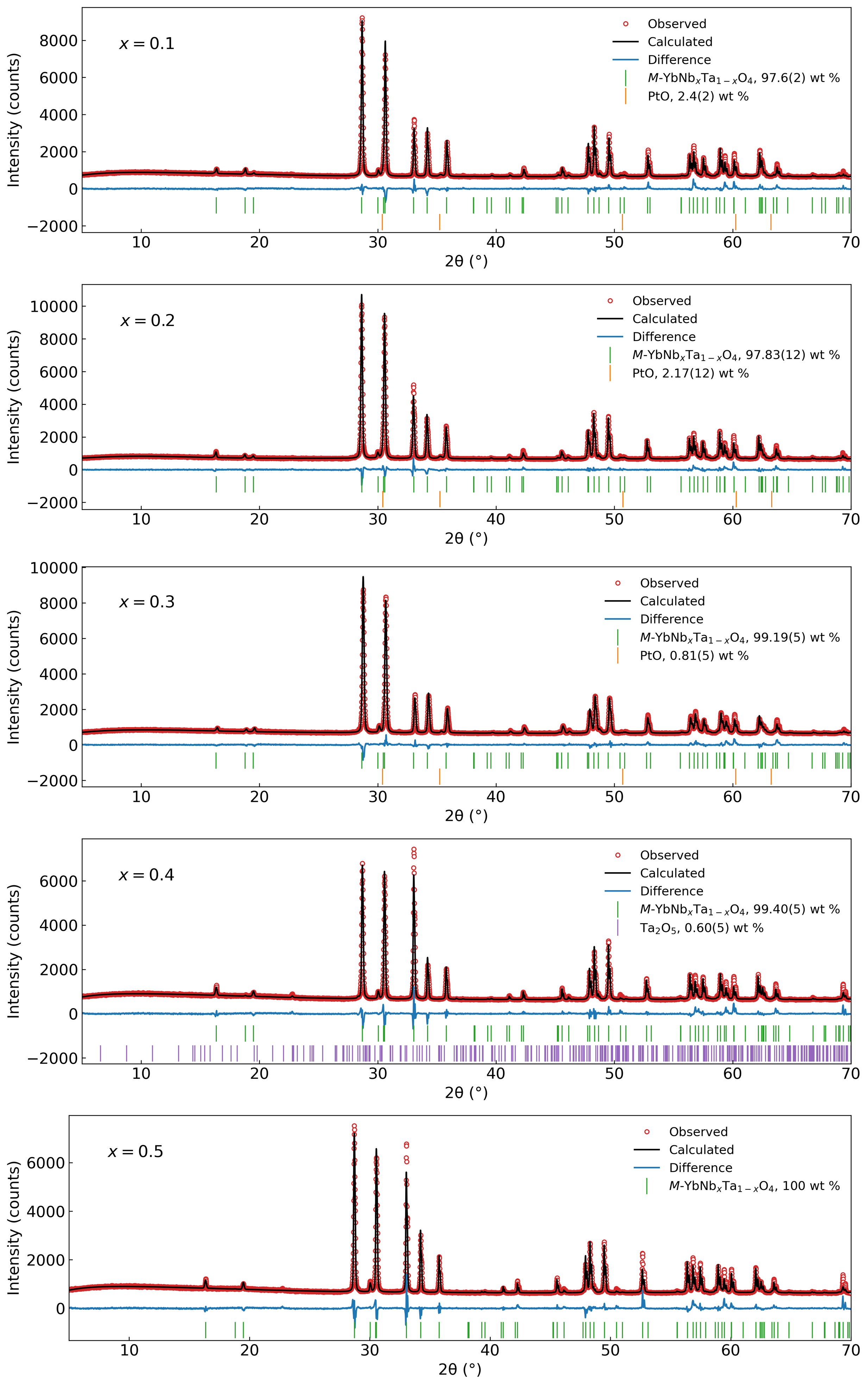}
\caption{Rietveld refinements against room-temperature PXRD data ($\lambda=$ Cu K$\alpha$) for \ch{YbNb_{x}Ta_{1-x}O4} with $x=0.1, 0.2, 0.3, 0.4, 0.5$ (top to bottom), synthesised under HPHT conditions (6~GPa, 1800~\degree C, 1~h). Red circles: observed data, black line: calculated pattern, blue line: difference pattern, tick marks: Bragg reflection positions.}
\label{fig:rietveld1to5}
\end{figure}

\begin{figure}[htbp]
\centering
\includegraphics[width=0.75\textwidth]{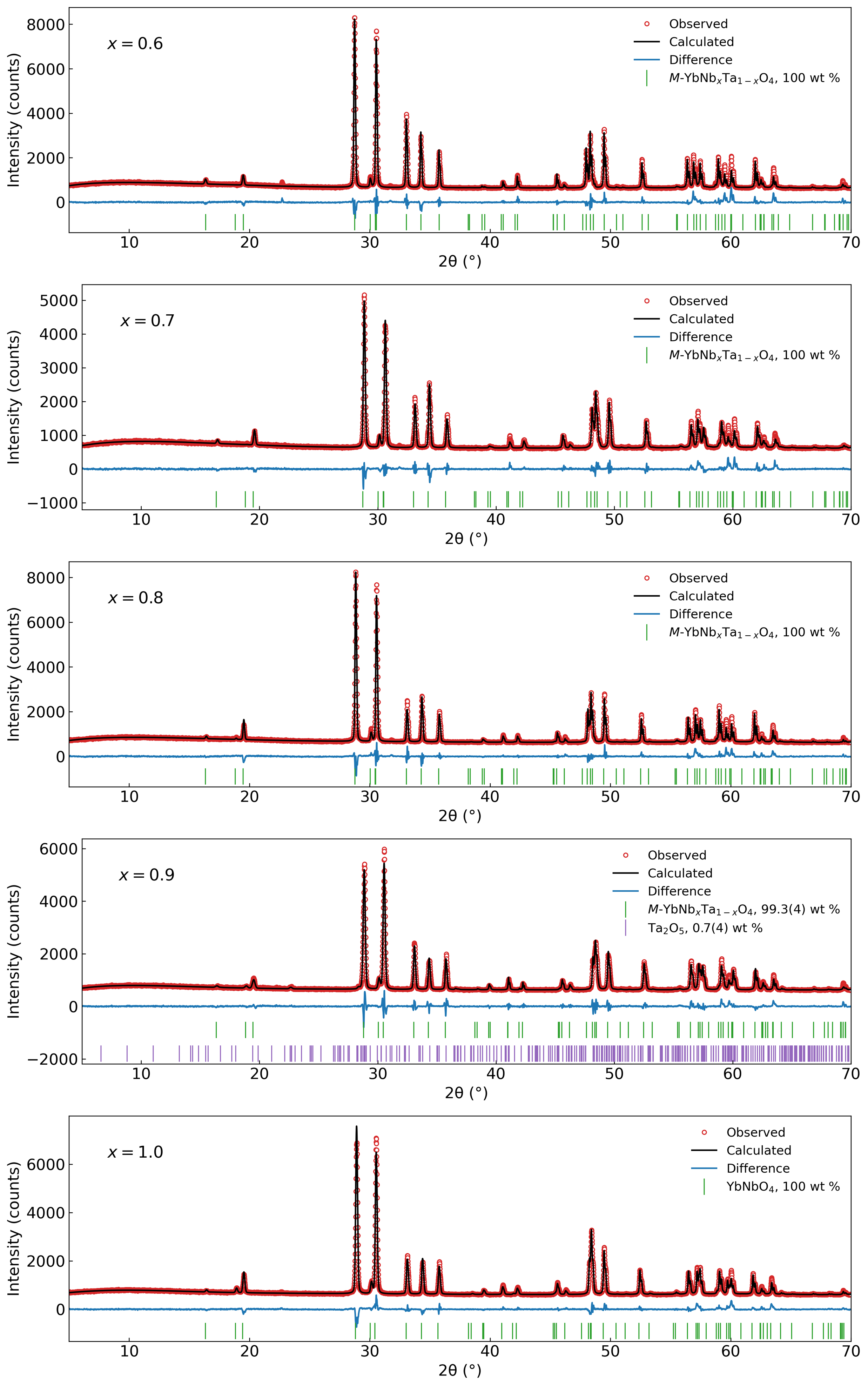}
\caption{Rietveld refinements against room-temperature PXRD data ($\lambda=$ Cu K$\alpha$) for \ch{YbNb_{x}Ta_{1-x}O4} with $x=0.6, 0.7, 0.8, 0.9, 1.0$ (top to bottom), synthesised under HPHT conditions (6~GPa, 1800~\degree C, 1~h), as-made. Red circles: observed data, black line: calculated pattern, blue line: difference pattern, tick marks: Bragg reflection positions.}
\label{fig:rietveld6to10}
\end{figure}

\begin{figure}[htbp]
\centering
\includegraphics[width=\textwidth]{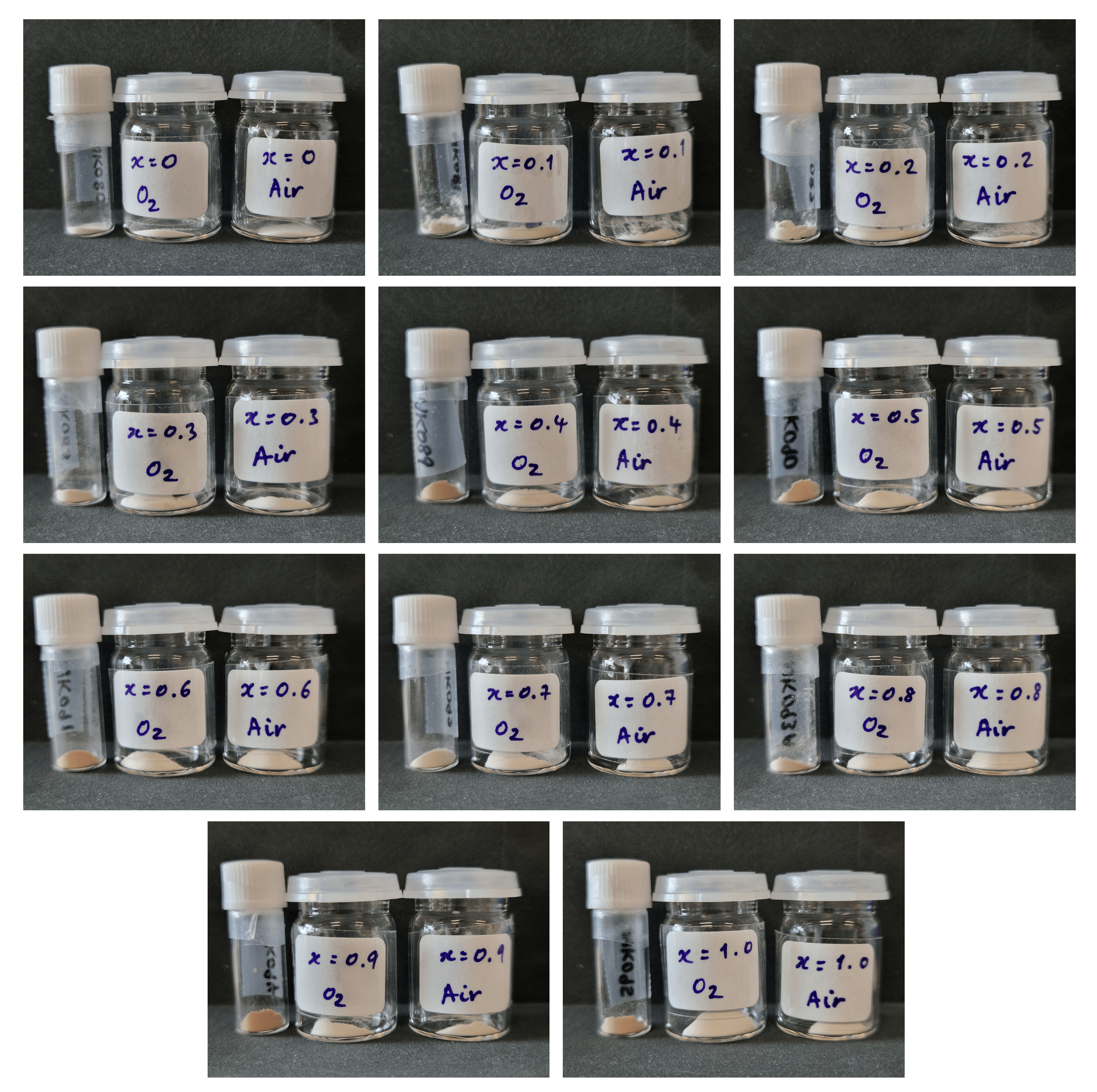}
\caption{Images showing the colour change from off-white or beige to pure white in \ch{YbNb_{x}Ta_{1-x}O4} samples after annealing at 800~\degree C for 2~h.}
\label{fig:annealedphotos}
\end{figure}

\begin{figure}[htbp]
\centering
\includegraphics[width=0.5\textwidth]{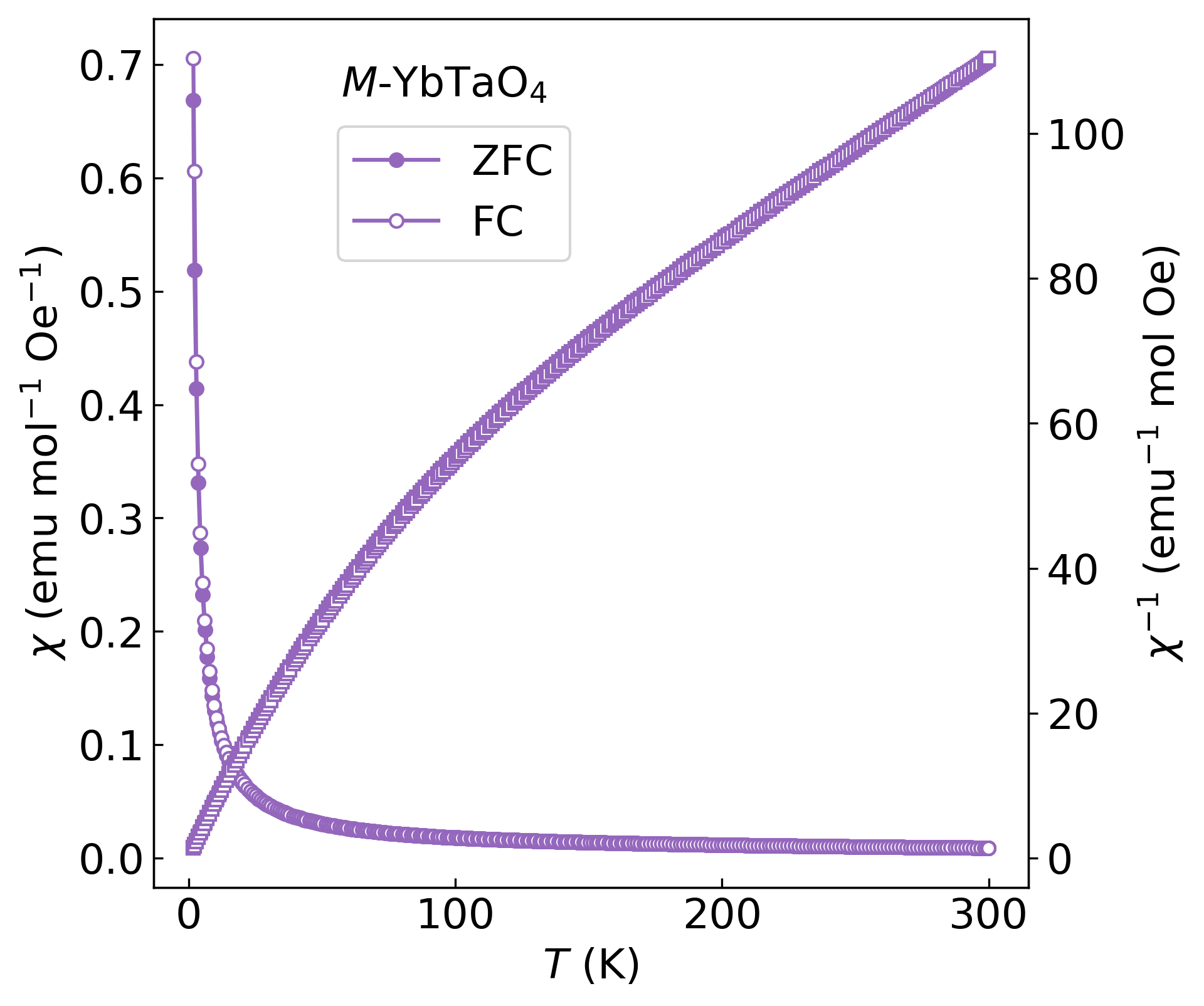}
\caption{Magnetic susceptibility $\chi(T)$ and inverse susceptibility $\chi^{-1}(T)$ for a sample of $M$-\ch{YbTaO4} (NK080). The zero-field-cooled (ZFC) and field-cooled (FC) curves match, showing that there is no glassiness.}
\label{fig:no_spin_glass}
\end{figure}

\begin{table}[htbp]
\centering
\caption{Fitted Curie-Weiss parameters for $M$-\ch{YbTaO4} samples using the two-level Curie-Weiss function (Eq.~3 in main text).}
\label{table:ybtao4_two_level_function_allsamples}
\begin{tabular}{l | c c c c}
\toprule
Sample & $\mu_0$ ($\mu_\mathrm{B}$) & $\mu_1$ ($\mu_\mathrm{B}$) & $\Delta E_{10}$ (K)& $\theta_\mathrm{CW}$ (K) \\
\midrule
NK071  &  3.55 & 5.91 & 262  & --2.5 \\
NK078  &  3.49 & 5.73 & 265  & --2.8 \\
NK079  &  3.53 & 5.78 & 233  & --1.7 \\
NK080  &  3.62 & 6.70 & 287  & --3.8 \\
NK096  &  3.55 & 5.90 & 247  & --2.0 \\
NK097  &  3.41 & 5.58 & 241  & --2.2 \\
NK098A &  3.57 & 5.61 & 275  & --1.6 \\
NK099  &  3.49 & 5.62 & 264  & --2.0 \\
NK100  &  3.40 & 5.92 & 262  & --3.1 \\
NK101  &  3.39 & 5.74 & 248  & --2.3 \\
NK102  &  3.37 & 5.56 & 234  & --2.0 \\ 
\bottomrule
\end{tabular}
\end{table}

\begin{table}[htbp]
\centering
\caption{Fitted Curie-Weiss parameters for $M$-\ch{YbTaO4} samples using the simple Curie-Weiss law, $\chi=C/(T-\theta_\mathrm{CW})$, in the high- and low-temperature regimes.}
\label{table:ybtao4_simple_cw_allsamples}
\begin{tabular}{l | c c c|c c c}
\toprule
Sample & Range (K) & $\mu_\mathrm{eff}$ ($\mu_\mathrm{B}$) & $\theta_\mathrm{CW}$ (K) & Range (K) & $\mu_\mathrm{eff}^\mathrm{LT}$ ($\mu_\mathrm{B}$) & $\theta_\mathrm{CW}^\mathrm{LT}$ (K) \\
\midrule
NK071  & 150--300   & 4.96 & --87  & 1.8--30  & 3.37 & --0.63 \\
NK078  & 150--300   & 4.82 & --86  & 1.8--30  & 3.31 & --0.87 \\
NK079  & 150--300   & 4.83 & --70  & 1.8--30  & 3.42 & --0.74 \\
NK080  &  150--300  & 5.61 & --135 & 1.8--30   &3.35 & --0.88 \\
NK096  & 150--300   & 4.94 & --79  & 1.8--30  & 3.40 & --0.63 \\
NK097  & 150--300   & 4.68 & --75  & 1.8--30  & 3.26 & --0.66 \\
NK098A & 150--300   & 4.71 & --71  & 1.8--30  & 3.43 & --0.36 \\
NK099  & 150--300   & 4.72 & --77  & 1.8--30  & 3.34 & --0.57 \\
NK100  &  150--300  & 4.95 & --101 & 1.8--30   &3.20 & --0.83 \\
NK101  & 150--300   & 4.80 & --86  & 1.8--30  & 3.22 & --0.65 \\
NK102  & 150--300   & 4.65 & --74  & 1.8--30  & 3.23 & --0.68 \\ 
\bottomrule
\end{tabular}
\end{table}

\begin{figure}[htbp]
\centering
\includegraphics[width=0.75\textwidth]{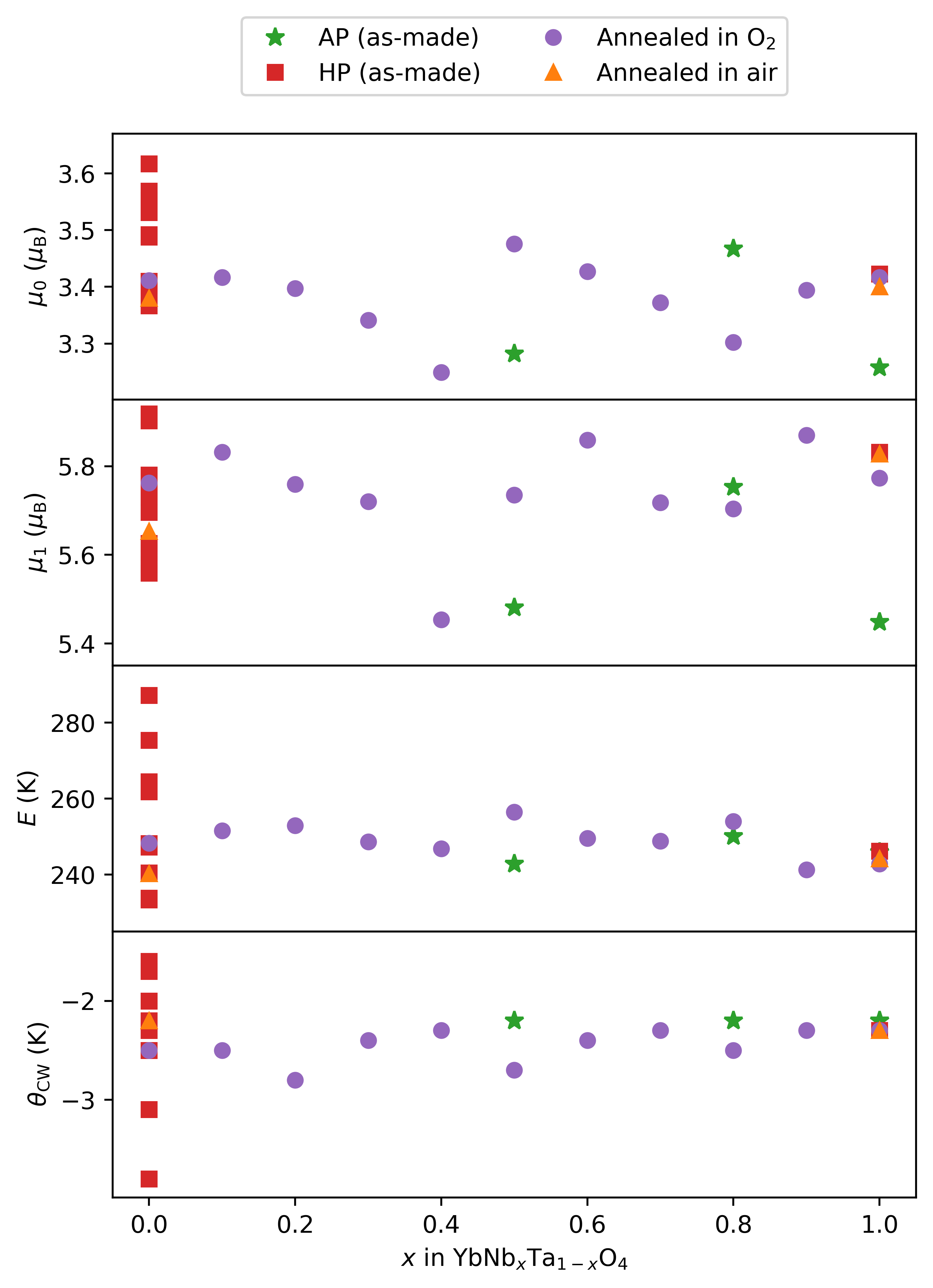}
\caption{Fitted magnetic parameters (two-level Curie-Weiss fitting) for selected samples of \ch{YbNb_{x}Ta_{1-x}O4}.}
\label{fig:NbTa_magnetic}
\end{figure}

\end{document}